\documentclass{llncs}
\usepackage[english]{babel}
\usepackage[latin1]{inputenc}
\usepackage{amsmath}
\usepackage{hhline}
\usepackage{amssymb}
\usepackage{latexsym}
\usepackage{xspace}
\usepackage{fancybox}
\usepackage{multicol}
\usepackage{pstricks}
\usepackage{pst-node}
\def\ignore#1{}
\ignore{ \setlength{\oddsidemargin}{0.25in}
\setlength{\textwidth}{6in} \setlength{\topmargin}{-0.25in}
\setlength{\textheight}{8.5in} }

\newcommand{\restr}[2]{{#1}|_{#2}}

\def\calS0{{\cal S}_0}

\def\cald{{\cal D}}
\def\calq{{\cal Q}}

\def\wnnw#1{\omega_{#1}}

\def\sat#1{{\it Sat}({#1})}

\ignore{
\newtheorem{definition}{Definition}
\newtheorem{theorem}{Theorem}
\newtheorem{corollary}{Corollary}
\newtheorem{proposition}{Proposition}

\newtheorem{example}{Example}

\newenvironment{proof}{\noindent{\em Proof.}}{\hfill\eop}
}
\title{\bf Semantic Optimization of Preference Queries
\thanks{Submitted. Research supported by
NSF Grant IIS-0307434.}}

\author{Jan Chomicki}

\institute{Dept. of Computer Science and Engineering\\ University
at Buffalo\\ Buffalo, NY 14260-2000\\ {\tt
chomicki@cse.buffalo.edu}}

\newcommand{\eop}{\hbox{\hskip6pt\vrule height 6pt width 6pt}}
\begin{document}
\maketitle
\begin{abstract}
Preference queries are relational algebra or SQL queries that
contain occurrences of the winnow operator ({\em find the most
preferred tuples in a given relation}).
 We present here a number of semantic optimization
techniques applicable to preference queries. The techniques make
it possible to remove redundant occurrences of the winnow operator
and to apply a more efficient algorithm for the computation of
winnow. We also study the propagation of integrity constraints in
the result of the winnow. We have identified necessary and
sufficient conditions for the applicability of our techniques, and
formulated those conditions as {\em constraint satisfiability}
problems.
\end{abstract}
\section{Introduction}\label{sec: introduction}
The notion of {\em preference} is becoming more and more
ubiquitous in present-day information systems. Preferences are
primarily used to filter and personalize the information reaching
the users of such systems. In database systems, preferences are
usually captured as {\em preference relations} that are used to
build {\em preference queries}
\cite{ChEDBT02,ChTODS03,Kie02,KiKo02}. From a formal point of
view, preference relations are simply binary relations defined on
query answers. Such relations provide an abstract, generic way to
talk about a variety of concepts like priority, importance,
relevance, timeliness, reliability etc. Preference relations can
be defined using logical formulas \cite{ChEDBT02,ChTODS03} or
special preference constructors \cite{Kie02} (preference
constructors can be expressed using logical formulas). The
embedding of preference relations into relational query languages
is typically provided through a relational operator that selects
from its argument relation the set of the {\em most preferred
tuples}, according to a given preference relation. This operator
has been variously called {\em winnow} (the term we use here)
\cite{ChEDBT02,ChTODS03}, BMO \cite{Kie02}, and Best
\cite{ToCi02}. (It is also implicit in skyline queries
\cite{BoKoSt01}.) Being a relational operator, winnow can clearly
be combined with other relational operators, in order to express
complex preference queries.

\begin{example}\label{ex:book}
We introduce an example used throughout the paper.
Consider the relation $Book(ISBN,Vendor,Price)$
and the following preference relation $\succ_{C_1}$ between {\em Book} tuples:
\begin{quote}
{\em prefer one Book tuple to another if and only if their ISBNs are the same and the Price of the
first is lower.}
\ignore{
{\em if two tuples have the same ISBN and different Price, prefer the one with the lower Price}.}
\end{quote}
Consider the instance $r_1$ of $Book$ in Figure \ref{fig:book}.
Then the winnow operator $\wnnw{C_1}$ returns the set of tuples in
Figure \ref{fig:winnow}.

\begin{figure}[htb]
\centering
\begin{tabular}{|l|l|l|}
\hline
{\em ISBN} &{\em Vendor} &{\em Price}\\\hline
0679726691 & BooksForLess & \$14.75\\
0679726691 & LowestPrices & \$13.50\\
0679726691 & QualityBooks & \$18.80\\
0062059041 & BooksForLess & \$7.30\\
0374164770 & LowestPrices & \$21.88\\\hline
\end{tabular}
\caption{The Book relation}
\label{fig:book}
\end{figure}

\begin{figure}[htb]
\centering
\begin{tabular}{|l|l|l|}
\hline
{\em ISBN} &{\em Vendor} &{\em Price}\\\hline
0679726691 & LowestPrices & \$13.50\\
0062059041 & BooksForLess & \$7.30\\
0374164770 & LowestPrices & \$21.88\\\hline
\end{tabular}
\caption{The result of winnow}
\label{fig:winnow}
\end{figure}
\end{example}
\begin{example}
The above example is a one-dimensional skyline query.
To see an example of a two-dimensional skyline, consider
the schema of {\em Book\/} expanded by another attribute {\em Rating\/}.
Define the following preference relation $C_2$:
\begin{quote}
{\em prefer one Book tuple to another if and only if their ISBNs are the same and the Price of the
first is lower and the Rating of the first is not lower, or the Price of the first is not higher
and the Rating of the first is higher.}
\end{quote}
Then $\wnnw{C_2}$ is equivalent to the following skyline (in the terminology of \cite{BoKoSt01}):
\begin{verbatim}
    SKYLINE ISBN DIFF, Price MIN, Rating MAX.
\end{verbatim}
The above notation indicates that only books with the same ISBN
should be compared, that Price should be minimized, and Rating
maximized. In fact, the tuples in the skyline satisfy the property
of {\em Pareto-optimality}, well known in economics.
\end{example}

Preference queries can be reformulated in relational algebra or SQL, and thus
optimized and evaluated using standard relational techniques. However, it has
been recognized that specialized evaluation and optimization techniques promise
in this context performance improvements that are otherwise unavailable.
A number of new algorithms for the evaluation of skyline queries (a special class
of preference queries) have been proposed \cite{BoKoSt01,ChGoGrLi03,KoRaRo02,PaTaFuSe03}.
Some of them can be used to evaluate general preference queries \cite{ChTODS03}.
Also, algebraic laws that characterize the interaction of winnow with the
standard operators of relational algebra have been formulated \cite{ChTODS03,KiHa02,KiHa03}.
Such laws provide a foundation for the rewriting of preference queries.
For instance, necessary and sufficient conditions for pushing a selection through
winnow are described in \cite{ChTODS03}.
The algebraic laws cannot be applied unconditionally.
In fact, the preconditions of their applications refer to the {\em validity}
of certain {\em constraint formulas}.

In this paper, we pursue the line of research from \cite{ChTODS03}
a bit further. We study {\em semantic optimization} of preference
queries. Semantic query optimization has been extensively studied
for relational and deductive databases \cite{CHGrMi90}. As a
result, a body of techniques dealing with specific query
transformations like join elimination and introduction, predicate
introduction etc. has been developed. We view semantic query
optimization very broadly and classify as {\em semantic} any query
optimization technique that makes use of integrity constraints. In
the context of preference queries, we focus on  the winnow
operator. Despite the presence of specialized evaluation
techniques, winnow is still quite an expensive operation. We
develop optimizing techniques that:
\begin{enumerate}
\item remove redundant occurrences of winnow;
\item recognize when more efficient evaluation of winnow is possible.
\end{enumerate}
More efficient evaluation of winnow can be achieved, for example,
if the given preference relation is a {\em weak order} (a
negatively transitive strict partial order). We show that even
when the preference relation is not a weak order (as in Example
\ref{ex:book}), it may become equivalent to a weak order on the
relations satisfying certain integrity constraints. We show a very
simple, single-pass algorithm for evaluating winnow under those
conditions. We also pay attention to the issue of satisfaction of
integrity constraints in the result of applying winnow. In fact,
some constraints may hold in the result of winnow, even though
they do not hold in the relation to which winnow is applied.
Combined with known results about the preservation of integrity
constraints by relational algebra operators \cite{Klu80,KlPr82},
our results provide a way for optimizing not only single
occurrences of winnow but also complex preference queries. As in
the case of the algebraic transformations described in
\cite{ChTODS03}, the semantic transformations described in this
paper have preconditions referring to the validity of certain
constraint formulas. Thus, such preconditions can be checked using
well established constraint satisfaction techniques
\cite{GuSuWe96}\footnote{A formula is valid iff its negation is
unsatisfiable.}.

The plan of the paper is as follows. In Section \ref{sec:basic} we
define basic notions. We limit ourselves here to integrity
constraints that are {\em functional dependencies}. In Section
\ref{sec:redundant} we address the issue of eliminating redundant
occurrences of winnow. In Section \ref{sec:weak} we study weak
orders. In Section \ref{sec:propagation} we characterize
dependencies holding in the result of winnow. In Section
\ref{sec:cgd} we show how our results can be generalized to
constraint-generating dependencies \cite{BaChWo99}. We briefly
discuss related work in Section \ref{sec:related} and conclude in
Section \ref{sec:concl}.

\section{Basic notions}\label{sec:basic}

We are working in the context of the relational model of data.
For concreteness, we consider two infinite domains: $\cald$ (uninterpreted constants) and $\calq$ (rational numbers).
Other domains could be considered as well without influencing most of the results of the paper.
We assume that database instances are finite.
Additionally,
we have the standard built-in predicates.

\subsection{Preference relations}
\begin{definition}\label{def:prefrel}
Given a relation schema $R(A_1 \cdots A_k)$
such that $U_i$, $1\leq i\leq k$, is the domain (either $\cald$ or $\calq$)
of the attribute $A_i$, a relation $\succ$ is a {\em preference relation over $R$}
if it is a subset of $(U_1\times\cdots\times U_k)\times (U_1\times\cdots\times U_k)$.
\end{definition}

Intuitively, $\succ$ will be a binary relation between tuples from the
same (database) relation.
We say that a tuple $t_1$ {\em dominates} a tuple $t_2$
in $\succ$ if $t_1\succ t_2$.

Typical properties of the relation $\succ$ include:
\begin{itemize}
\item {\em irreflexivity}: $\forall x.\ x\not\succ x,$
\item {\em asymmetry}: $\forall x,y.\ x\succ y\Rightarrow y\not\succ x,$
\item {\em transitivity}: $\forall x,y,z.\ (x\succ y \wedge y\succ z)\Rightarrow x\succ z,$
\item {\em negative transitivity}: $\forall x,y,z.\ (x\not\succ y \wedge y\not\succ z)\Rightarrow x\not\succ z,$
\item {\em connectivity}: $\forall x,y.\ x\succ y\vee y\succ x \vee x=y.$
\end{itemize}

The relation $\succ$ is:
\begin{itemize}
\item a {\em strict partial order} if it is
irreflexive and transitive (thus also asymmetric);
\item a {\em weak order} if it is a
negatively transitive strict partial order;
\item a {\em total order} if it is
a connected strict partial order.
\end{itemize}

At this point, we do not assume any properties
of $\succ$, although in most applications it will satisfy at least the properties of {\em strict partial
order}.

\begin{definition}\label{def:prefformula}
A {\em preference formula (pf)} $C(t_1,t_2)$ is a first-order
formula defining a preference relation $\succ_C$ in the standard
sense, namely
\[t_1\succ_C t_2\;{\rm iff}\; C(t_1,t_2).\]
An {\em intrinsic preference formula (ipf)} is a preference formula that
uses only built-in predicates.
\end{definition}

We will limit our attention to preference relations defined using intrinsic preference
formulas.

Because we consider two specific domains, $\cald$ and $\calq$, we will have two
kinds of variables, $\cald$-variables and $\calq$-variables, and two kinds
of atomic formulas:
\begin{itemize}
\item {\em equality  constraints}: $x=y$, $x\not=y$, $x=c$, or $x\not= c$, where
$x$ and $y$ are $\cald$-variables, and $c$ is an uninterpreted constant;
\item {\em rational-order constraints}: $x\theta y$ or $x\theta c$, where
$\theta\in\{=,\not=,<,>,\leq,\geq\}$, $x$ and $y$ are $\calq$-variables, and $c$ is a
rational number.
\end{itemize}

Without loss of generality, we
will assume that ipfs are in DNF (Disjunctive Normal Form) and
quantifier-free (the theories involving the above domains admit
quantifier elimination).
We also assume that atomic formulas are closed under negation (also
satisfied by the above theories).
An ipf whose all atomic formulas are equality  (resp. rational-order)
constraints will be called an {\em equality} (resp. {\em rational-order}) ipf.
Clearly, ipfs are a special case of general constraints
\cite{CDB00}, and define {\em fixed}, although possibly infinite,
relations.
By using the notation $\succ_C$ for a preference relation, we assume that there is
an underlying preference formula $C$.

\begin{definition}
Given an instance $r$ of $R$ and a preference relation $\succ_C$ over $R$,
the {\em restriction} $\restr{\succ_C}{r}$ of $\succ_C$ to $r$ is defined
as
\[\restr{\succ_C}{r}=\succ_C\ \cap\ r\times r.\]
\end{definition}

\subsection{Winnow}
We define now an algebraic operator that picks from a given relation the
set of the {\em most preferred tuples}, according to a given preference formula.
\begin{definition}\label{def:winnow}
If $R$ is a relation schema and $C$ a preference
formula defining a preference relation $\succ_C$ over $R$,
then the {\em winnow operator} is written as $\wnnw{C}(R)$,
and for every instance $r$ of $R$:
\[\wnnw{C}(r)=\{t\in r\mid\neg \exists t'\in r.\ t'\succ_C t\}.\]
\end{definition}

A preference query is a relational algebra query containing at least
one occurrence of the winnow operator.

\begin{example}\label{ex:book:02}
Consider the relation $Book(ISBN,Vendor,Price)$ (Example
\ref{ex:book}). The preference relation $\succ_{C_1}$ from this
example can be defined using the formula $C_1$:
\[(i,v,p)\succ_{C_1}(i',v',p') \equiv i=i' \wedge p<p'.\]
The answer to the preference query
$\wnnw{C_1}(Book)$
provides for every book the information about the vendors offering the
lowest price for that book.
\end{example}
\subsection{Indifference}
Every preference relation $\succ_C$ generates an indifference relation $\sim_C$:
two tuples $t_1$ and $t_2$ are {\em indifferent}
($t_1\sim_C t_2$) if
neither is preferred to the other one, i.e.,
$t_1\not\succ_C t_2$ and $t_2\not\succ_C t_1$.

\begin{proposition}\label{prop:indiff}
For every preference relation $\succ_C$, every relation $r$ and every tuple $t_1,t_2\in\wnnw{C}(r)$,
we have $t_1=t_2$ or $t_1\sim_C t_2$.
\end{proposition}

\subsection{Functional dependencies}
We assume that we are working in the context of a single relation
schema and all the integrity constraints are over that schema. The
set of all instances of $R$ satisfying a set of integrity
constraints $F$ is denoted as $\sat{F}$. We say that $F$ {\em
entails} an integrity constraint $f$ if every instance satisfying
$F$ also satisfies $f$.

A functional dependency (FD) $f\equiv X\rightarrow Y$, where $X$ and $Y$ are sets
of attributes of $R$ can be written down as the following logic formula:
\[\forall t_1.\forall t_2.\ [R(t_1)\wedge R(t_2)\wedge t_1[X]=t_2[X]] \Rightarrow
t_1[Y]=t_2[Y].\] We use the following notation:
\[\varphi_f(t_1,t_2)\equiv t_1[X]=t_2[X] \Rightarrow t_1[Y]=t_2[Y].\]
For a set of FDs  $F$, we define
\[\varphi_F\equiv\bigwedge_{f\in F}\varphi_f.\]

The {\em arity} of an FD $f\equiv X\rightarrow Y$ is the cardinality $|X\cup Y|$ of the set of
attributes $X\cup Y$.
The {\em arity} of a set of FDs $F$ is the maximum arity of any FD in $F$.

Note that the set of attributes $X$ in $X\rightarrow Y$ may be empty, meaning
that each attribute in $Y$ can assume only a single value.

\section{Eliminating redundant occurrences of
winnow}\label{sec:redundant}

Given an instance $r$ of $R$, the operator $\wnnw{C}$ is redundant if $\wnnw{C}(r)=r$.
If we consider the class of all instances of $R$, then such an operator is redundant
for every instance iff $\succ_C$ is an empty relation.
The latter holds iff $C$ is unsatisfiable.
However, we are interested only in the instances satisfying a given set of integrity
constraints.
Therefore, we will check whether the restriction $\restr{\succ_C}{r}$ is empty
for every instance $r$ satisfying the given set of integrity constraints.

\begin{definition}
Given a set of integrity constraints $F$, the operator $\wnnw{C}$ is {\em redundant
w.r.t. a set of integrity constraints $F$} if $\forall r\in\sat{F}$, $\wnnw{C}(r)=r$.
\end{definition}

\begin{theorem}\label{th:redundant}
$\wnnw{C}$ is redundant w.r.t. a set of FDs $F$ iff the following formula is unsatisfiable:
\[\varphi_F(t_1,t_2)\wedge t_1\succ_C t_2\]
\end{theorem}
\begin{proof}
Assume that formula in the theorem is satisfiable.
Then there are tuples $t_a$ and $t_b$ such that $\varphi_F(t_a,t_b)$ and $t_a\succ_C t_b$.
Thus $t_b\not\in\wnnw{C}(\{t_a,t_b\})$ and thus $\wnnw{C}$ is not redundant w.r.t. $F$.
For the other direction, assume $\wnnw{C}$ is not redundant w.r.t. $F$.
Then there is an instance $r_0\in\sat{F}$ and a tuple $t_b\in r_0$ such that
$t_b\not\in \wnnw{C}(r_0)$. Thus, there must be a tuple $t_a$ in $r_0$ such that
$t_a\succ_c t_b$. Clearly, $\varphi_F(t_a,t_b)$ and therefore the formula in the theorem
is satisfiable.
\end{proof}

Theorem \ref{th:redundant}
shows that checking for redundancy w.r.t. a set of FDs $F$ is a {\em constraint satisfiability} problem.

\begin{example}\label{ex:book:03}
Consider Example \ref{ex:book:02} in which the FD $ISBN\rightarrow Price$ holds. Then
\[\varphi_F\equiv i_1=i_2\Rightarrow p_1=p_2\]
and $\varphi_F(t_1,t_2)\wedge t_1\succ_{C_1}t_2$ is
\[(i_1=i_2\Rightarrow p_1=p_2)\wedge i_1=i_2\wedge p_1<p_2
.\] The last formula is clearly unsatisfiable, and thus the
implication in Theorem \ref{th:redundant} holds and we can infer
that $\wnnw{C_1}$ is redundant w.r.t. $ISBN\rightarrow Price$.
\end{example}

How hard is it to check for redundancy w.r.t. a set of FDs $F$?
We assume that the size of a preference
formula $C$ (over a relation $R$) in DNF
is characterized by two parameters: ${\it width\/}(C)$ -- the number of disjuncts in $C$,
and ${\it span\/}(C)$ -- the maximum number of conjuncts in a disjunct of $C$.
Namely, if $C=D_1\vee\cdots\vee D_m$, and each $D_i=C_{i,1}\wedge\cdots C_{i,k_i}$,
then ${\it width\/}(C)=m$ and ${\it span\/}(C)=\max \{k_1,\ldots,k_m\}$.

\begin{theorem}\label{th:check}
If:
\begin{itemize}
\item the cardinality of the set of FDs $F$ is $|F|$ and its arity is at most $k$;
\item the given preference relation is defined using an ipf $C$ containing only atomic constraints over the same domain
and such that ${\it width\/}(C)\leq m$, ${\it span\/}(C)\leq n$;
\item the time complexity of checking satisfiability of a conjunctive ipf with $n$ conjuncts is in $O(T(n))$,
\end{itemize}
then the time complexity of checking $\wnnw{C}$ for redundancy
with respect to  $F$ is in $O(m\ k^{k |F|}\ T(\max(k |F|,n)))$.

\end{theorem}

The paper \cite{GuSuWe96} contains several results about checking satisfiability
of conjunctive formulas. For instance, in the case of rational-order formulas,
this problem is shown to be solvable in $O(n)$. This implies, for example,
the following corollary.

\begin{corollary}\label{cor:check}
If a preference relation is  defined by a conjunctive
rational-order ipf ($m=1$) and the arity of $F$ is at most $2$,
then checking $\wnnw{C}$ for redundancy w.r.t. $F$ can be done in
time $O(n\ 2^{|F|})$ .
\end{corollary}
An analogous result can be derived for equality formulas. From now on
we will only present detailed complexity analysis for rational-order
formulas.

\section{Weak orders}\label{sec:weak}

We have defined weak orders as negatively transitive strict partial orders.
Equivalently, they can be defined as strict partial orders for which the
indifference relation is transitive.
Intuitively, a weak order consists of a number (perhaps infinite) of linearly ordered layers.
In each layer, all the elements are mutually indifferent and they are all above all the elements in lower
layers.
\begin{example}\label{ex:book:04}
In the preference relation $\succ_{C_1}$ in Example \ref{ex:book:02}, the first, second and third tuples are
indifferent with the fourth and fifth tuples. However, the first tuple is preferred to the
second, violating the transitivity of indifference. Therefore, the preference relation $\succ_{C_1}$
is not a weak order.
\end{example}
\begin{example}
A preference relation $\succ_{C_f}$, defined as
\[ x\succ_{C_f} y \equiv f(x)>f(y)\]
for some real-valued function $f$, is a weak order but not a total
order.
\end{example}

\subsection{Computing winnow}
Many algorithms for evaluating winnow are possible. However, we
discuss here those that have a good {\em blocking} behavior and
thus are capable of processing very large data sets.

We first review BNL (Figure \ref{fig:BNL}), a basic algorithm for evaluating winnow, and show
that for preference relations that are weak orders a much simpler and
more efficient algorithm is possible.
BNL was proposed in \cite{BoKoSt01} in the context
of {\em skyline queries}. However, \cite{BoKoSt01} also noted
that the algorithm requires only the properties of strict partial orders.
BNL uses  a fixed amount of main memory (a {\em window}).
It also needs a temporary table for the tuples whose status cannot be
determined in the current pass, because the available amount of main memory is limited.
\begin{figure}[htb]
\centering
\fbox{%
\begin{minipage}{.8\textwidth}
\begin{small}\begin{enumerate}
\item clear the window $W$ and the temporary table $F$;
\item make $r$ the input;
\item repeat the following until the input is empty:
\begin{enumerate}
\item for every tuple $t$ in the input:
\begin{itemize}
\item $t$ is dominated by a tuple in $W$ $\Rightarrow$ ignore $t$,
\item $t$ dominates some tuples in $W$ $\Rightarrow$ eliminate the dominated tuples
and insert $t$ into $W$,
\item if $t$ and all tuples in $W$ are mutually indifferent $\Rightarrow$ insert $t$ into $W$
(if there is room), otherwise add $t$ to $F$;
\end{itemize}
\item output the tuples from $W$ that were added there when $F$ was empty,
\item make $F$ the input, clear the temporary table.
\end{enumerate}
\end{enumerate}
\caption{BNL: Blocked Nested Loops}
\label{fig:BNL}
\end{small}\end{minipage}}
\end{figure}

BNL keeps in the window the best tuples discovered so far (some of
them may also be in the temporary table). All the tuples in the
window are mutually indifferent and they all need to be kept,
since each may turn out to dominate some input tuple arriving
later. For weak orders, however, if a tuple $t_1$ dominates $t_2$,
then any tuple indifferent to $t_1$ will also dominate $t_2$. In
this case, indifference is an equivalence relation, and thus it is
enough to keep in main memory only a single tuple $top$ from the
top equivalence class. In addition, one has to keep track of all
members of that class (called the {\em current bucket} $B$), since
they may have to be returned as the result of the winnow. The new
algorithm WWO (Winnow for Weak Orders) is shown in Figure
\ref{fig:WWO}.

\begin{figure}[htb]
\centering
\fbox{%
\begin{minipage}{.8\textwidth}
\begin{small}\begin{enumerate}
\item $top$ := the first input tuple
\item $B:=\{top\}$
\item for every subsequent tuple $t$ in the input:
\begin{itemize}
\item $t$ is dominated by $top$ $\Rightarrow$ ignore $t$,
\item $t$ dominates $top$ $\Rightarrow$ $top:=t$; $B:=\{t\}$
\item $t$ and $top$ are indifferent $\Rightarrow$ $B:=B\cup\{t\}$
\end{itemize}
\item output $B$
\end{enumerate}
\caption{WWO: Weak Order Winnow} \label{fig:WWO}
\end{small}\end{minipage}}
\end{figure}

It is clear that WWO requires only a single pass over the input.
It uses additional memory (whose size is at most equal to the size
of the input) to keep track of the current bucket. However, this
memory is only written and read once, the latter at the end of the
execution of the algorithm. Clearly, for weak orders WWO is
considerably more efficient than BNL. Note that for weak orders
BNL does not simply reduce to WWO. Note also that if additional
memory is not available, WWO can execute in a small, fixed amount
of memory by using two passes over the input: in the first, a top
tuple is identified, and in the second, all the tuples indifferent
to it are selected.

 In \cite{ChGoGrLi03} we
proposed SFS, a more efficient variant of BNL for skyline queries,
in which a presorting step is used. Because sorting may require
more than one pass over the input, that approach will also be less
efficient than WWO for weak orders.

\subsection{Relative weak orders}

Even if a preference relation $\succ_C$ is not a weak order in
general, its restriction to a specific instance or a class of
instances may be a weak order, and thus WWO may be applied to the
computation of winnow. Again, we are going to consider the class
of instances $\sat{F}$ for a set of integrity constraints $F$.

\begin{definition}\label{def:relative}
A preference relation $\succ_C$ is a {\em weak order relative to a set
of integrity constraints $F$} if $\forall r\in\sat{F}$, $\restr{\succ_C}{r}$
is a weak order.
\end{definition}

\begin{theorem}\label{th:relative}
An irreflexive preference relation $\succ_C$ is a weak order relative to a set of
FDs $F$ iff the following formula is unsatisfiable:
\[\varphi_F(t_1,t_2)\wedge\varphi_F(t_2,t_3)\wedge\varphi_F(t_1,t_3)\wedge
t_1\succ_C t_2\wedge t_1\sim_C t_3\wedge t_2\sim_C t_3.\]
\end{theorem}
\begin{example}\label{ex:book:05}
Consider Example \ref{ex:book:02}, this time with the $0$-ary FD
$\emptyset \Rightarrow ISBN$. (Such a dependency might hold, for
example, in a relation resulting from the selection $\sigma_{ISBN=c}$
for some constant $c$.)
Note that
\[(i,v,p)\sim_c (i',v',p') \equiv i\not=i' \vee p=p'.\]
We construct the following formula, according to Theorem \ref{th:relative}:
\[i_1=i_2\wedge i_2=i_3\wedge i_1=i_3\wedge i_1=i_2\wedge p_1<p_2 \wedge
(i_1\not= i_3\vee p_1= p_3) \wedge (i_2\not= i_3\vee p_2= p_3)\]
which is unsatisfiable. Therefore, $\succ_{C_1}$ is a weak order
relative to the FD $\emptyset \Rightarrow ISBN$, and for every
instance $r$ satisfying this dependency, $\wnnw{C_1}(r)$ can be
computed using the single-pass algorithm WWO.
\end{example}
\begin{theorem}\label{th:wo}
If:
\begin{itemize}
\item the cardinality of the set of FDs $F$ is $|F|$ and its arity is at most $k$;
\item the given preference relation is defined using an ipf $C$ containing only atomic constraints over the same domain
and such that ${\it width\/}(C)\leq m$, ${\it span\/}(C)\leq n$;
\item the time complexity of checking satisfiability of a conjunctive ipf with $n$ conjuncts is in $O(T(n))$,
\end{itemize}
then the time complexity of checking whether $\succ_C$ is a weak
order relative to  $F$ is in $O(m\ n^{4m}\ k^{k |F|}\ T(\max(k
|F|,m,n)))$.
\end{theorem}

\begin{corollary}\label{cor:wo}
If a preference relation is  defined by a conjunctive
rational-order ipf ($m=1$) and the arity of $F$ is at most $2$,
then then the time complexity of checking whether $\succ_C$ is a
weak order relative to  $F$ is in $O( n^{5}\ 2^{|F|})$.
\end{corollary}
\section{Propagation of integrity
constraints}\label{sec:propagation}

The study of propagation of integrity constraints by relational
operators is essential for semantic optimization of complex
queries. We need to know which integrity constraints hold in the
results of such operators. The winnow operator returns a subset of
a given relation, thus it preserves all the functional
dependencies holding in the relation. However, we also know that
winnow returns a set of tuples which are mutually indifferent.
This property can be used to derive {\em new} dependencies that
hold in the result of winnow without necessarily holding in the
input relation. (New dependencies can also be derived for other
relational operators, for example selection, as in Example
\ref{ex:book:05}.)

\begin{theorem}\label{th:new}
Let $f$ be an FD and $\succ_C$ an irreflexive preference relation
over $R$. The following formula
\[t_1\sim_C t_2 \wedge \neg\varphi_f(t_1,t_2)\]
is unsatisfiable iff for every instance $r$ of $R$,
$\wnnw{C}(r)$ satisfies $f$.
\end{theorem}
\begin{proof}
We will call the FDs satisfying the condition in Theorem
\ref{th:new} {\em generated} by $\succ_C$ and denote the set of
all such dependencies by $G_C$. It is easy to show that $G_C$ is
closed w.r.t. FD implication. Assume $f\not\in G_C$. Then the
formula in the theorem is satisfiable. Assume it is satisfied by
tuples $t_a$ and $t_b$ ($t_a\not= t_b$ because otherwise
$\neg\varphi(t_a,t_b)$ is false). Thus
$r_0=\{t_a,t_b\}\not\in\sat{f}$. But $t_a\sim_C t_b$,
$t_a\not\succ_{C} t_a$, and $t_b\not\succ_{C} t_b$. Thus
$r_0=\wnnw{C}(r_0)\not\in\sat{f}$.

In the other direction, assume that there is an instance $r_0$
such that $\wnnw{C}(r_0)\not\in\sat{f}$. By the properties of FDs,
we can assume that $\wnnw{C}(r_0)$ consists of two distinct tuples
$t_a$ and $t_b$. By Proposition \ref{prop:indiff}, we know that
$t_a\sim_C t_b$. Thus the formula is satisfied by $t_a$ and $t_b$.
\end{proof}

\begin{example}\label{ex:book:06}
Consider Example \ref{ex:book:02}.
Then the formula from Theorem \ref{th:new} is
\[(i_1\not=i_2 \vee p_1=p_2)\wedge i_1=i_2\wedge p_1\not=p_2\]
which is clearly unsatisfiable.
Thus, the FD~ $ISBN\rightarrow Price$ holds in the result of $\wnnw{C_1}$,
even though it might not hold in the input relation.
\end{example}
\begin{theorem}\label{th:generation}
If:
\begin{itemize}
\item the arity of $f$ is $k$;
\item the given preference relation is defined using an ipf $C$ containing only atomic constraints over the same domain
and such that ${\it width\/}(C)\leq m$, ${\it span\/}(C)\leq n$;
\item the time complexity of checking satisfiability of a conjunctive ipf with $n$ conjuncts is in $O(T(n))$,
\end{itemize}
then the time complexity of checking checking the condition in
Theorem \ref{th:new} is in $O(k n^{2m}\ T(\max(k,m)))$.
\end{theorem}
\begin{corollary}\label{cor:generation}
If a preference relation is  defined by a conjunctive
rational-order ipf ($m=1$) and the arity of $f$ is at most $2$,
then the time complexity of checking the condition in Theorem
\ref{th:new} is in $O( n^{2})$.
\end{corollary}
\section{Constraint-generating dependencies}\label{sec:cgd}

Functional dependencies are a special case of {\em
constraint-generating dependencies\/} \cite{BaChWo99}.
\begin{definition}\label{def:cgd}
A {\em constraint-generating dependency (CGD)} can be expressed a formula of the following form:
\[\forall t_1.\ldots\forall t_n.\ [R(t_1)\wedge\cdots\wedge R(t_n)\wedge \gamma(t_1,\ldots t_n)]
\Rightarrow \gamma'(t_1,\ldots t_n)\] where $\gamma(t_1,\ldots
t_n)$ and $\gamma'(t_1,\ldots t_n)$ are constraints over some
constraint theory.
\end{definition}
CGDs are equivalent to denial constraints.
\begin{example}
We give here some examples of CGDs. Consider the relation {\it
Emp} with attributes {\it Name}, {\it Salary}, and {\it Manager},
with {\it Name} being the primary key. The constraint that {\it no
employee can have a salary greater that that of her manager} is a
CGD:
\[\forall n,s,m,s',m'.~[{\it Emp\/}(n,s,m)\wedge{\it Emp\/}(m,s',m')]\Rightarrow
s\leq s'.\]
Similarly, single-tuple constraints ({\tt CHECK} constraints in SQL2) are
a special case of CGDs. For example, the constraint that
{\em no employee can have a salary over \$200000} is expressed as:
\[\forall n,s,m.~{\it Emp\/}(n,s,m)\Rightarrow s\leq 200000].\]
\end{example}

It turns out that the problems studied in the present paper can be
viewed as specific instances of the {\em entailment} (implication)
of CGDs. To see that, let's define two special CGDs $d^{\;C}_2$
and $d^{\;C}_3$ for a given preference relation $\succ_C$ (and the
corresponding indifference relation $\sim_C$):
\[d^{\;C}_2\equiv\forall t_1.\forall t_2.\ R(t_1)\wedge R(t_2)\Rightarrow t_1\sim_C t_2\]
and
\[d^{\;C}_3\equiv\forall t_1.\forall t_2.\forall t_3.\ \ R(t_1)\wedge R(t_2)\wedge R(t_3)\Rightarrow
\neg(t_1\succ_C t_2\wedge t_1\sim_C t_3\wedge t_2\sim_C t_3).\]
Then we have the following properties that generalize Theorems \ref{th:redundant}, \ref{th:relative}, and \ref{th:new}.
\begin{theorem}\label{th:cgd:redundant}
$\wnnw{C}$ is redundant w.r.t. a set of CGDs $F$ iff $F$ entails
$d^{\;C}_2$.
\end{theorem}
\begin{theorem}\label{th:cgd:weak}
If $\succ_C$ is irreflexive, then $\succ_C$ is a weak order
relative to a set of CGDs $F$ iff $F$ entails $d^{\;C}_3$.
\end{theorem}
\begin{theorem}\label{th:cgd:new}
If $\succ_C$ is irreflexive,
then a CGD $f$ is entailed by $d^{\;C}_2$ iff for every instance
$r$ of $R$, $\wnnw{C}(r)$ satisfies $f$.
\end{theorem}

\begin{example}
Consider the following preference relation $\succ_{C_{\alpha}}$
where $\alpha$ is a selection condition over the schema $R$:
\[t_1\succ_{C_{\alpha}}t_2 \equiv
\alpha(t_1)\wedge\neg\alpha(t_2).\] This is a very common
preference relation expressing the preference for the tuples
satisfying some property over those that do not satisfy it. The
corresponding indifference relation $\sim_{{C_{\alpha}}}$is
defined as follows:
\[t_1\sim_{C_{\alpha}} t_2 \equiv
\alpha(t_1)\wedge\alpha(t_2)\vee\neg\alpha(t_1)\wedge\neg\alpha(t_2).\]
Theorem \ref{th:cgd:redundant} implies that $\wnnw{C_{\alpha}}$ is
redundant w.r.t. a set of CGDs $F$ iff $F$ implies the CGD
\[\forall t_1.\forall t_2.\ R(t_1)\wedge R(t_2)\Rightarrow
\alpha(t_1)\wedge\alpha(t_2)\vee\neg\alpha(t_1)\wedge\neg\alpha(t_2).\]
The latter dependency is satisfied by an instance $r$ of $R$ if
and only if all the tuples in $r$ satisfy $\alpha$ or none does.
In both cases $\wnnw{C_{\alpha}}(r)=r$.
\end{example}

 The paper \cite{BaChWo99}
contains an effective reduction using {\em symmetrization} from
entailment of CGDs to  validity of  $\forall$-formulas in the
underlying constraint theory. (A similar construction using {\em
symbol mappings} is presented in \cite{ZhOz97}.) This immediately
gives the decidability  of the problems discussed in the present
paper for equality and rational-order constraints (as well as other
constraint theories for which  satisfiablity of quantifier-free formulas is decidable).
A more detailed complexity analysis can be carried out
along the lines of Theorems \ref{th:check}, \ref{th:wo}, and
\ref{th:generation}.

For theorems \ref{th:cgd:redundant},\ref{th:cgd:weak} and
\ref{th:cgd:new} to hold for a class of integrity constraints, two
conditions need to be satisfied: (a) the class should be able to
express constraints equivalent to $d^{\;C}_2$ and $d^{\;C}_3$ ,
and (b) the notions of entailment and finite entailment (entailment
on finite relations) for the
class should coincide. If (b) is not satisfied,
then the theorems will still hold if  reformulated by replacing
"entailment" with "finite entailment". Thus, assuming that
(a) is satisfied, the effectiveness of checking the preconditions
of the above theorems depends on the decidability of finite
entailment for the given class of integrity constraints.

\section{Related work}\label{sec:related}
The basic reference for semantic query optimization is
\cite{CHGrMi90}. The most common techniques are: join
elimination/introduction, predicate elimination and introduction, and
detecting an empty answer set. \cite{GryzVLDB99} discusses the
implementation of predicate introduction and join elimination in
an industrial query optimizer. Semantic query optimization
techniques for relational queries are studied in \cite{ZhOz97}
in the context of denial and referential constraints,
and in \cite{MaWa00} in the context of constraint
tuple-generating dependencies (a generalization of CGDs and classical
relational dependencies).
FDs are used for reasoning about
sort orders in \cite{SiShMa96}.

Two different approaches to preference queries have been pursued
in the literature: qualitative and quantitative. In the {\em
qualitative} approach, represented by
\cite{LaLa87,KiGu94,KoKiThGu95,BoKoSt01,GoJaMa01,ChEDBT02,ChTODS03,Kie02,KiHa02,KiKo02},
the preferences between tuples in the answer to a query are
specified directly, typically using binary {\em preference
relations}. In the  {\em quantitative} approach
\cite{AgWi00,HrKoPa01}, preferences are specified indirectly using
{\em scoring functions} that associate a numeric score with every
tuple of the query answer. Then a tuple $t_1$ is preferred to a
tuple $t_2$ iff the score of $t_1$ is higher than the score of
$t_2$. The qualitative approach is strictly more general than the
quantitative one, since one can define preference relations in
terms of scoring functions However, not every intuitively
plausible preference relation can be captured by scoring
functions.
\begin{example}\label{ex:book:01}
There is no scoring
function that captures the preference relation described in Example \ref{ex:book}.
Since there is no preference defined between any of the first three tuples
and the fourth one, the score of the fourth tuple should be equal to all of
the scores of the first three tuples. But this implies that the scores
of the first three tuples are the same, which is not possible since
the second tuple is preferred to the first one which in turn is preferred
to the third one.
\end{example}
This lack of expressiveness of the quantitative approach is well known
in utility theory \cite{Fish99,Fish70}.
The importance of weak orders in this context comes from the fact that only weak orders
can be represented using real-valued scoring functions (and for countable domains
this is also a sufficient condition for the existence of such a representation \cite{Fish70}).
In the present paper we do not assume that preference relations are weak orders.
We only characterize a condition under which preference relations become weak orders
relative to a set of integrity constraints.

Algebraic optimization of preference queries is discussed in \cite{ChTODS03,KiHa02,KiHa03}.

\section{Conclusions and further work}\label{sec:concl}
We have presented some techniques for semantic optimization of
preference queries, focusing on the winnow operator. The
simplicity of our results attests to the power of logical
formulation of preference relations. However, our results are
applicable not only to the original logical framework of
\cite{ChEDBT02,ChTODS03}, but also to preference queries defined
using preference constructors \cite{Kie02,KiKo02} and skyline
queries \cite{BoKoSt01,ChGoGrLi03,KoRaRo02,PaTaFuSe03} because
those queries can be expressed using preference formulas.

Further work can address, for example, the following issues:
\begin{itemize}
\item identifying other semantic optimization
techniques for preference queries,
\item expanding the class of integrity constraints by considering, e.g., tuple-generating
dependencies and referential integrity constraints,
\item identifying weaker but easier to check sufficient conditions for the application
of our techniques,
\item considering other preference-related operators like {\em ranking} \cite{ChTODS03}.
\end{itemize}
\newcommand{\etalchar}[1]{$^{#1}$}

\end{document}